\documentclass[%
 aip,
 amsmath,amssymb,
 reprint,%
]{revtex4-1}

\usepackage{graphicx}
\usepackage{dcolumn}
\usepackage{bm}

\usepackage[utf8]{inputenc}
\usepackage[T1]{fontenc}
\usepackage{mathptmx}
\usepackage{etoolbox}
\usepackage{booktabs}

\makeatletter
\def\@email#1#2{%
 \endgroup
 \patchcmd{\titleblock@produce}
  {\frontmatter@RRAPformat}
  {\frontmatter@RRAPformat{\produce@RRAP{*#1\href{mailto:#2}{#2}}}\frontmatter@RRAPformat}
  {}{}
}%
\makeatother
\begin{document}

\preprint{AIP/123-QED}

\title{Absolute charge calibration of DRZ phosphor screens for relativistic electron bunches}

\author{Shuang Liu}
\affiliation{Institute of High Energy Physics, Chinese Academy of Sciences, Beijing 100049, China}

\author{Bing Zhou}
\affiliation{Laboratory of Zhongyuan Light, School of Physics, Zhengzhou University, Zhengzhou, 450001, China}
\affiliation{Department of Engineering Physics, Tsinghua University, Beijing 100084, China}

\author{Pengwei Huang}
\affiliation{Department of Engineering Physics, Tsinghua University, Beijing 100084, China}

\author{Tianliang Zhang}
\affiliation{Department of Engineering Physics, Tsinghua University, Beijing 100084, China}

\author{Bo Peng}
\affiliation{Laboratory of Zhongyuan Light, School of Physics, Zhengzhou University, Zhengzhou, 450001, China}

\author{Shengtai Wei}
\affiliation{Department of Engineering Physics, Tsinghua University, Beijing 100084, China}

\author{Zhiyuan Guo}
\affiliation{Department of Engineering Physics, Tsinghua University, Beijing 100084, China}

\author{Bo Guo}
\affiliation{Beijing Academy of Quantum Information Sciences, Beijing 100193, China}

\author{Fei Li}
\affiliation{Institute of High Energy Physics, Chinese Academy of Sciences, Beijing 100049, China}

\author{Jianfei Hua}
\affiliation{Department of Engineering Physics, Tsinghua University, Beijing 100084, China}

\author{Yang Wan}
\email[]{yangwan23@zzu.edu.cn}
\affiliation{Laboratory of Zhongyuan Light, School of Physics, Zhengzhou University, Zhengzhou, 450001, China}

\author{Wei Lu}
\email[]{weilu@ihep.ac.cn}
\affiliation{Institute of High Energy Physics, Chinese Academy of Sciences, Beijing 100049, China}
\affiliation{Beijing Academy of Quantum Information Sciences, Beijing 100193, China}


\begin{abstract}
Laser-plasma accelerators have been the subject of extensive research in recent years. The electron beams they generate exhibit a broad energy spread. To conveniently characterize beams from laser wakefield acceleration (LWFA), electron spectrometers employing scintillating screens coupled with CCD cameras are typically used. In this work, we calibrate a series of DRZ phosphor screens and measure the spectra of the light they emit. The calibration was performed using the radio-frequency linear electron accelerator at Tsinghua University, which provided monoenergetic electron beams with peak energy of approximately 30 MeV.
\end{abstract}

\maketitle

\section{Introduction}

Laser wakefield acceleration has attracted great interest worldwide over past years with the development of high-power femtosecond laser system. In 1979, Tajima and Dawson proposed the principle that accelerating electrons in the wakefield excited by a laser pulse in plasma\cite{PhysRevLett.43.267}. In a typical LWFA, an ultrashort laser pulse is focused onto a gas jet. The pulse front is strong enough to ionize neutral gas to plasma, and the main part of the pulse drives nonlinear wakefield. The accelerating gradient can reach 100 GV/m, about three orders of magnitude higher compared to radio-frequency accelerators, capable of generating quasimonoenergetic electron bunches \cite{2004A,2004High,2004Monoenergetic} approaching 10 GeV within only a centimeter scale distance\cite{2006GeV,XiaomingWang2013Quasi,picksley2024matched}. The LWFA beam has quite different parameters from the bunches delivered by LINAC, such as a high charge density \cite{2017Demonstration}, a short duration\cite{2011Few,zhang2016temporal} and small source sizes\cite{weingartner2012ultralow,wan2023femtosecond,laberge2024revealing,huang2024electro}, promising for various applications such as compact radiotherapy \cite{guo2025preclinical,glinec2006radiotherapy,labate2020toward,Svendsen2021LPA-VHEE}, ultrafast imaging \cite{wan2025femtosecond,wan2022direct,wan2023femtosecond,wan2024real,zhang2017femtosecond,levine2025direct}, advanced light sources \cite{wang2021free,labat2023seeded,pompili2022free,barber2025greater}. For a typical LWFA, the energy distribution of the accelerated electron bunch is usually continuous from several MeV to hundreds of MeV at repetition rates below 10 Hz with shot-to-shot fluctuations. Measuring the charge and energy variations in single-shot mode is thus crucial for studying and improving the stability of LWFAs.
In conventional radio-frequency (RF) accelerators, Faraday cups are typically employed to measure electrons with energies below 5 MeV. At higher electron beam energies, the escape of secondary electrons compromises measurement accuracy. While an Integrating Current Transformer (ICT) can accurately determine the charge of a collimated electron beam, its finite aperture limits its ability to measure the beam energy spectrum across a broad energy range when deflected by a dipole magnet.

Scintillating screens possess numerous advantages, including a short millisecond luminescence decay time, simple and low-cost fabrication, high luminescence efficiency, reusability, and have been widely used in the LWFA field for detecting and characterizing relativistic electron beams. These screens, typically composed of gadolinium oxysulfide doped with terbium ($\rm{Gd_2O_2S:Tb}$), convert the energy of incident electrons into visible light, enabling single-shot spatially resolved beam profiling and charge measurement via imaging systems. The accurate conversion from measured light intensity to incident charge density is paramount for quantitative analysis. However, as these screens are primarily marketed for X-ray applications, their absolute response to electron beams is not specified by manufacturers, necessitating rigorous calibration for precise charge determination.

The pioneering work by Glinec established an early absolute calibration for a Lanex screen using 3-9 MeV electrons. They reported a linear response and derived an intrinsic conversion efficiency of $\sim$16\% for the $\rm{Gd_2O_2S:Tb}$ phosphor, providing a foundational benchmark for the field\cite{2006Absolute}. This was followed by a comprehensive study by Buck et al., which calibrated eight different screens\cite{2010Absolute}. Their work systematically quantified the absolute light yield ($14.8\times10^9 \rm{ph/sr/pC}$ for Kodak BioMAX MS) and established a linear response of several orders of magnitude. Subsequent work by Kurz et al.,\cite{2018Calibration} found that the absolute calibration factors from prior work were approximately two times too high, implying an underestimation of the charge in the LPA experiments. Their measurements, extending to charge densities of 10 $\rm{nC/mm^2}$, detailed the saturation behavior using Birks' law and provided a systematic report on long-term degradation under irradiation. 

In addition to the Kodak series, the DRZ screen series that promises higher sensitivity have also been widely used in LWFAs, for which Wu et al., provided early data for DRZ-High and PI-200\cite{2012Note}, reporting yields of 7.82 and 12.01 ($×10^9 \rm{ph/sr/pC}$), respectively, and confirming their Lambertian emission, and Schwinkendorf et al., later calibrated three types of DRZ (High, Plus, Std)\cite{2019Charge}, showing that DRZ High offered a 30\% higher yield than Kodak BioMAX MS and demonstrated improved robustness. Most recently, Chiang et al. measured the absolute light yields of LANEX regular and PI200 to be 7.39 ($×10^9 \rm{ph/sr/pC}$) and 20.4 ($×10^9 \rm{ph/sr/pC}$), respectively \cite{Chiang2024Absolute}. However, these studies on DRZ screens were limited to specific subsets, resulting in fragmented and sometimes inconsistent data. A unified calibration of the entire DRZ family under identical conditions is still lacking, creating ambiguity for LWFA researchers. To establish a consistent and definitive reference for the laser-plasma accelerator community, this work was undertaken to perform a unified calibration of the prevalent DRZ screens under identical experimental conditions. A key aspect of our methodology involved the precise measurement of the absolute luminescence spectrum for each screen. This crucial step ensures that the transmission efficiency of our optical system and the response of the camera are accurately accounted for, thereby significantly enhancing the reliability of the absolute charge calibration we present.

\section{setup}
\begin{figure}
    \centering
    \includegraphics[width=1\linewidth]{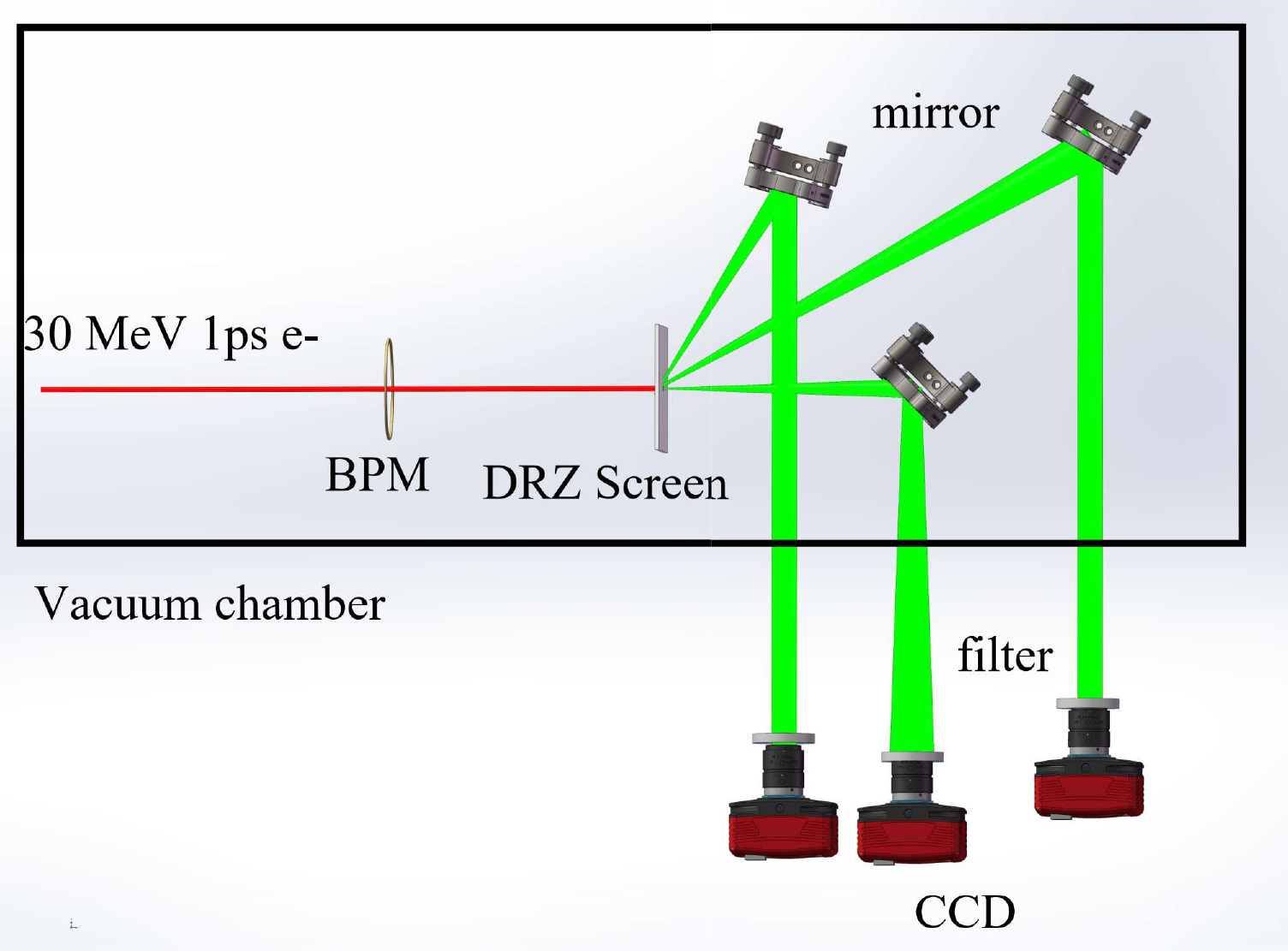}
    \caption{The experimental layout. electrons generated by a photocathode RF  gun accelerated to about 30 MeV through an accelerating tube, with energy spread of 1\%. The beam passed through the BPM and incident at the screen to be calibrated, while the BPM would calculate the charge at the same time.}
    \label{fig:2 }
\end{figure}

The experiment was performed using the radio frequency linear electron accelerator of the TTX platform in Tsinghua University, which accelerated the electron bunch to 30 MeV\cite{TANG2009Thomson}. A photocathode RF gun can generate the electron bunch with the maximum charge of 100 pC. After passing through an S band accelerating waveguide, the maximum energy can reach 30 MeV with the energy spread of 1\%. The bunch duration is about 1 ps. In our experiment, the beam charge of 30 MeV electron beams were tuned from 2 pC to 100 pC. A beam position monitor (BPM) calibrated by a Faraday cup was installed downstream the accelerating tube to measure the charge of the bunch. After passing through the BPM, 
the electron beams were transported to the scintillator screen and the energy deposed in the screen was converted to the photons with about 543 nm wavelength. The photons were collected by an optical imaging system, and finally coupled to a CCD. The imaging system composed of a metal coated mirror, an imaging lens and a band-pass filter with 546 nm central wavelength and 80 nm FWHM. The background noise including the dark current of the accelerator were measured and subtracted when processing the measured light signals. A vacuum compatible motorized linear stage with a holder was adopted to switch the screen under irradiation, all of the DRZ series screen (DRZ HR, DRZ FINE, DRZ Standard, DRZ Plus, DRZ High and PI200) were calibrated in the experiment.

\section{Results}

\subsection{spectra of PI200}
\begin{figure}
    \centering
    \includegraphics[width=1\linewidth]{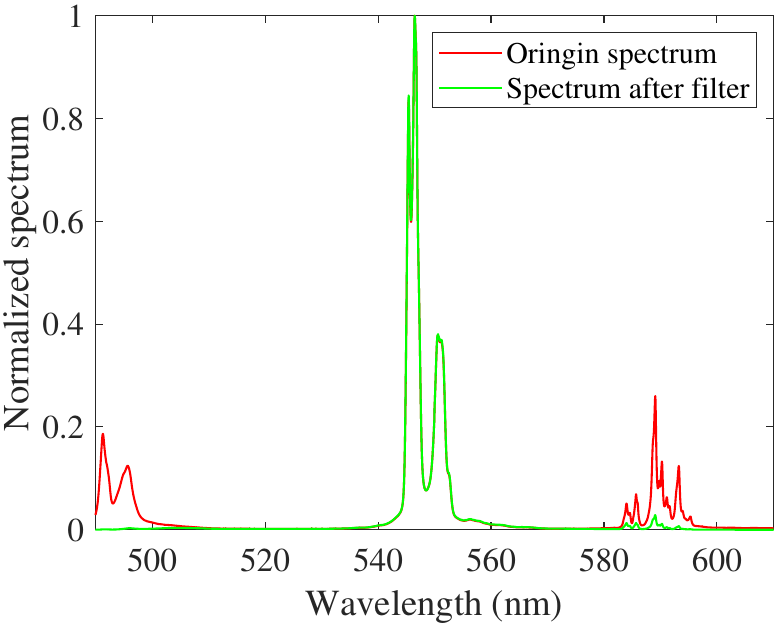}
    \caption{Emission Spectrum of PI200. The spectrum shows three main emission peaks at 489, 543, and 586 nm as red line. With the bandpass filter, peak at 543 was preserved and others were filtered that plot with green line.}
    \label{fig:2 }
\end{figure}

Prior to the absolute calibration of the scintillation screens, the emission photon spectra was measured using a imaging spectrometer. As shown in Figure 2, the photon emission of the $\rm{Gd_2O_2S:Tb}$ phosphor consists of 3 main emission bands in the measured spectra range centered at 489 nm, 543 nm, and 586 nm, respectively. 
Generally, the quantum efficiency of a CCD is not uniform over such a broad spectral range. To avoid ambiguity, we employed a band-pass filter (Edmund Optics \#87-009, center wavelength: 546 nm, bandwidth: 80 nm) to isolate the dominant green emission peak. Nevertheless, a small fraction of residual blue and yellow emission remained unfiltered. In the subsequent calculation of luminescence efficiency, the contributions from these residual components were subtracted based on the measured relative intensity, thereby yielding the accurate efficiency of the primary green emission peak. In the following sections, parameters and quantities related to photon counting, such as luminescence efficiency, will specifically refer to photon emission around the main peak at 543 nm.

\subsection{Angular distribution }

\begin{figure}
    \centering
    \includegraphics[width=1\linewidth]{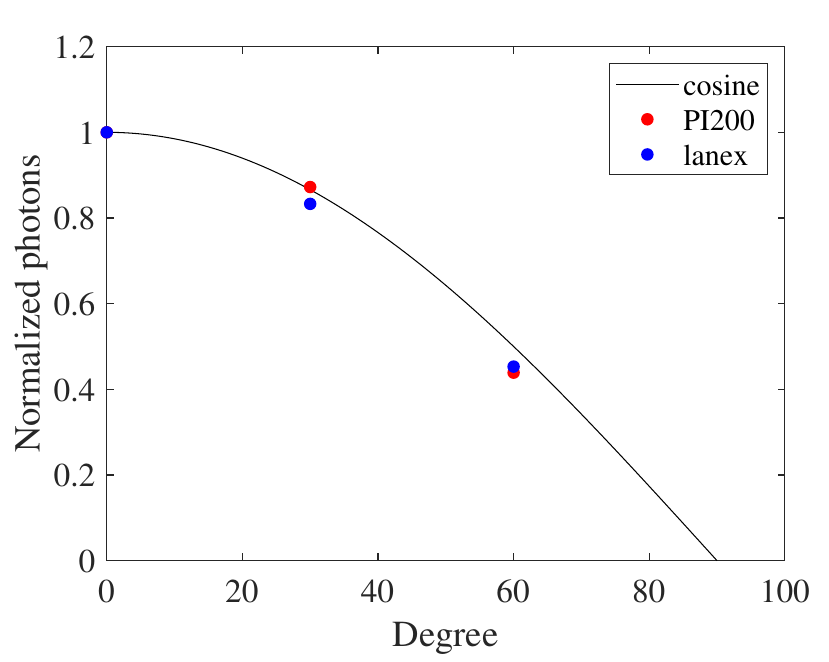}
    \caption{Angular distribution of emission photons of PI200 and lanex. Three observation angles  were chosen to measured the photons emitted. The two kinds of screen both shows the cosine variation}
    \label{fig:3 }
\end{figure}

To accurately convert the light signal captured by our off-axis imaging system into the total photons emitted by the scintillator, a precise understanding of its angular emission distribution is essential. For this purpose, we characterized the angular dependence of the light yield by employing a multi-camera setup. As illustrated in Figure 1, three identical CCD cameras were positioned at different angles relative to the normal surface of the screen. This configuration allowed for simultaneous acquisition of the scintillation signal at multiple angles under identical electron beam conditions, thereby eliminating potential shot-to-shot fluctuations. The integrated intensity of each camera was normalized to the value obtained in the normal direction. The resulting data, plotted as normalized intensity versus observation angle, exhibits a close agreement with the theoretical cos(${\theta}$) curve, which describes a Lambertian (or cosine) emitter. This observation is consistent with the results reported in other experiments, confirming that the DRZ screens behave as Lambertian sources. Consequently, the total integrated radiance (photons/sr) measured by our primary diagnostic camera can be reliably related to the total flux emitted into the forward hemisphere using the known angular distribution.

\subsection{Absolute fluorescence efficiency }

\begin{figure}
    \centering
    \includegraphics[width=1\linewidth]{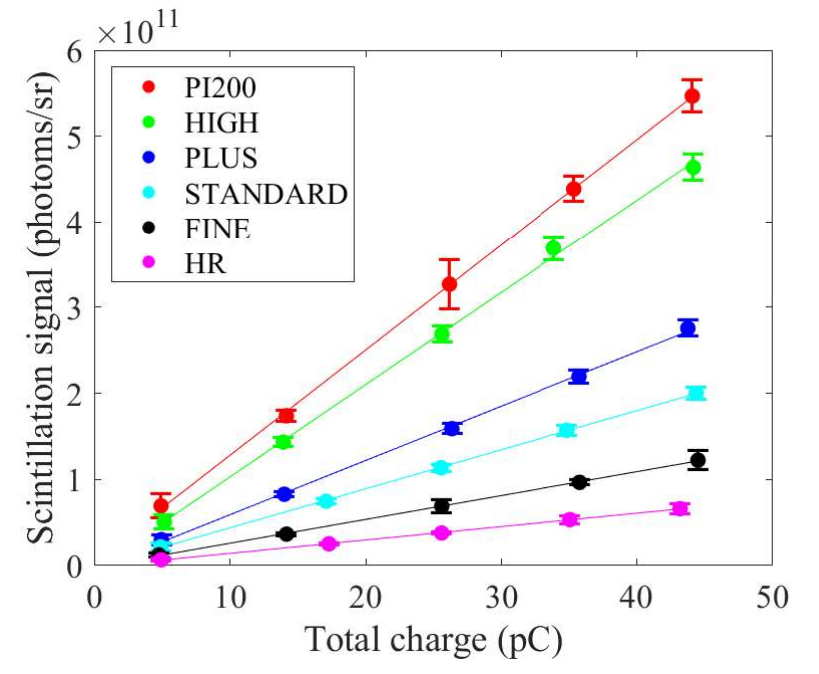}
    \caption{Absolute photons collected of all DRZ scintillation screen. charge varies from 5pC to 50 pC and scintillation signal varies linear, that shows the stability of screen for different charge}
    \label{fig:3 }
\end{figure}

\begin{table}[h]
    \centering
    \setlength{\tabcolsep}{20pt}
    \caption{Absolute fluorescence efficiency of DRZ screen. The most sensitive PI200 is up to 12.7$\times 10^9$ photons/sr/pC, and the lowest is only 1.5$\times 10^9$ photons/sr/pC. The errors consist of the charge quantity jitter of the electron bunches generated by the accelerator, inaccuracies in the optical system, and the standard deviation from raw images, calculated by combining these components in quadrature.}
    \label{tab:placeholder_label}
    \begin{tabular}{ccc}
        \toprule
        Name & Efficiency  & Relative error \\
             & ($10^9$ photons/sr/pC) &   \\
        \midrule
        PI200 & 12.7  & 5.9\% \\
        HIGH  & 10.4  & 6.6\% \\
        PLUS  & 6.4   & 5.0\% \\
        STD   & 4.5   & 4.9\% \\
        FINE  & 2.7   & 6.2\% \\
        HR    & 1.5   & 6.1\% \\
        \bottomrule
    \end{tabular}
\end{table}

To establish the absolute calibration curve and verify the linear response of the scintillation screens, the incident electron bunch charge was varied from 5 pC to 50 pC. The resulting scintillation light yield, quantified in photons per steradian, exhibited a highly linear dependence on the charge over this range, as confirmed by a linear least-squares fit to the data presented in Figure 4. This observed linearity is a critical prerequisite for a reliable charge diagnostic. The absolute calibration factor 
(in units of photons/sr/pC) for each screen type was subsequently determined from the slope of this linear fit.

\subsection{Error Analysis} 
A detailed uncertainty analysis was conducted to evaluate the precision of our absolute calibration factors, with the primary error budgets summarized in Table 1. The total measurement uncertainty originates from three principal, independent sources:
\begin{itemize}

\item Beam Charge Fluctuations: The intrinsic shot-to-shot charge jitter of the electron bunches delivered by the accelerator contributes an estimated uncertainty of approximately 2\%.
\item Optical System Efficiency: The combined uncertainty in the photon collection and detection efficiency arises from the tolerances of individual optical components. This includes the reflectivity of the mirror, the transmission of the vacuum window and lens, the precise attenuation of the ND filters, and the quantum efficiency of the CCD camera. These individual uncertainties were combined in quadrature, yielding a total systematic uncertainty for the optical path of approximately 5\%.
\item Image Signal Statistics: The statistical uncertainty was derived from the standard deviation of the integrated signal across a dataset of 100 consecutive images acquired under identical conditions, resulting in a contribution of about 2\%.
\end{itemize}
The total combined uncertainty for each calibration factor was finally computed by propagating these three independent components in quadrature.

\subsection{Discussion}

Prior to our experiments, three different research groups had conducted calibration studies on the absolute light yield efficiency of DRZ-series scintillator screens. In 2012, Wu et al. calibrated the light yield of the PI200 screen using a linear accelerator\cite{2012Note}. In 2019, J.-P. Schwinkendorf et al. calibrated three types of DRZ-series screens: DRZ High, DRZ Plus, and DRZ Standard\cite{2019Charge}. In 2024, Chen-Wei Chiang et al. calibrated the PI200 scintillator screen\cite{Chiang2024Absolute}. These results are summarized in Table \ref{tab:result_comparison}. It is evident that this work presents the first complete calibration of the entire DRZ series. Moreover, the absolute light yield measurements reported in the three previous studies exhibit considerable discrepancies. Although these differences may be attributed to variations in experimental details such as the choice of standard light sources and optical filters, they also introduce confusion. Our experiment provides an additional reference in this regard. Notably, our measurements for the DRZ Standard, DRZ Plus, and DRZ High screens are in close agreement with those reported in \cite{2019Charge}, demonstrating that consistent results can be achieved under precisely controlled experimental conditions. Ref \cite{Chiang2024Absolute} measured the full-spectrum light yield of the DRZ screens, which may explain the higher values compared to our results. Meanwhile, the bandpass filters used in \cite{2012Note} differ from those employed in our study, likely causing the observed discrepancies.

\begin{table}[h]
    \centering
    \setlength{\tabcolsep}{6pt}
    \caption{Comparison of the absolute light yield measurements between this work and \cite{2012Note,2019Charge,Chiang2024Absolute}, in unit of  $10^9 \rm{photons/sr/pC}$.}
    \begin{tabular}{ccccc}
    \toprule
    Screen Type & Wu et al. & J.-P. et al. & Chiang et al. & Our results \\ 
    \midrule
    HR    & --- & --- & --- & $1.5$ \\
    FINE  & --- & --- & --- & $2.7$ \\
    STD & --- & $5.5$ & --- & $4.5$ \\
    PLUS  & --- & $7.3$ & --- & 6.4\ \\
    HIGH  & $7.8$ & $10.6$ & --- & $10.4$ \\
    PI200 & $12.0$ & --- & $20.4$ & $12.7$ \\
    \bottomrule
    \end{tabular}
    \label{tab:result_comparison}
\end{table}

\section{Conclusion}
In this work, we measured the emission spectrum of the DRZ screen and calibrated the entire series of the DRZ screen. We characterized the emission spectrum of the DRZ screen, identifying distinct peaks at 489 nm, 543 nm, and 586 nm. The calibration focused specifically on the dominant 543 nm peak to determine its luminous efficiency. This methodology was subsequently applied to the entire DRZ screen series, enabling precise determination of their absolute luminous efficiencies.  The corresponding uncertainty values for these measurements, derived from a comprehensive error analysis, are systematically presented in Table 1. Furthermore, we confirmed that the spatial distribution of the emission follows a cosine dependence, which is consistent with previously reported work. 

\section{Acknowledgements}
This work was supported by the Strategic Priority Research Program of the Chinese Academy of Sciences (Grants No. XDB0530000), National Natural Science Foundation of China (Grants No.12574380), Science Fund Program for Distinguished Young Scholars of the National Natural Science Foundation of China (Overseas), Discipline Construction Foundation of ``Double World-class Project'', Key Scientific Research Projects of Henan Provincial Colleges and Universities No. 25ZX002, and Natural Science Foundation of Henan Province No.252300421300. 

\bibliography{ref}

\end{document}